\documentclass[useAMS,usenatbib]{mn2e}
\usepackage{epsfig}
\usepackage{natbib}
\usepackage{graphicx}

\title[An iterative filter to reconstruct planetary transits]{An
  iterative filter to reconstruct planetary transit signals in the
  presence of stellar variability}

\author[Alapini \& Aigrain]{A. Alapini$^{1}$\thanks{E-mail:
    alapini@astro.ex.ac.uk (AA); suz@astro.ex.ac.uk (SA)} and S.
  Aigrain$^{1}$\footnotemark[1]\\
  $^{1}$Astrophysics Group, School of Physics, University of Exeter,
  Stocker Road, Exeter, EX4 4QL, United Kingdom}

\begin{document}

\date{Accepted 2009 May 7.  Received 2009 May 6; in original form 2008 November 28}

\pagerange{\pageref{firstpage}--\pageref{lastpage}} \pubyear{pubyear}

\maketitle

\label{firstpage}

\begin{abstract}

  The detrending algorithms which are widely used to reduce the impact
  of stellar variability on space-based transit surveys are ill-suited
  for estimating the parameters of confirmed planets, as they
  unavoidably alter the transit signal. We present a post-detection
  detrending algorithm, which filters out signal on other timescales
  than the period of the transit while preserving the transit signal.

  We compare the performance of this new filter to a well-established
  pre-detection detrending algorithm, by applying both to a set of 20
  simulated light curves containing planetary transits, stellar
  variability, and instrumental noise as expected for the CoRoT space
  mission, and performing analytic fits to the transits. Compared to
  the pre-detection benchmark, the new post-detection filter
  systematically yields significantly reduced errors (median reduction
  in relative error over our sample $\sim 40\%$) on the planet-to-star radius
  ratio, system scale and impact parameter. This is particularly
  important for active stars, where errors induced by variability can
  otherwise dominate the final error budget on the planet parameters.

  Aside from improving planet parameter estimates, the new filter
  preserves all signal at the orbital period of the planet, and thus could
  also be used to search for light reflected by the planet.

\end{abstract}


\section{Introduction}\label{sec:intro}

Accurate measurements of the fundamental parameters of extra-solar
planets are needed to constrain theoretical models of planet formation
and evolution. For transiting planets, it is possible to measure the
radius and true mass, which can be confronted to the predictions of
evolutionary models with various compositions and heat deposition
mechanisms (e.g.\ \citealt{guil05,bcb08}). These models are
continuously challenged by new discoveries, the best known case of
this being the small group of planets whose radii are larger than
expected for their mass and irradiation level, such as HD\,209458b
\citep{cbl+00,kcb+07}, HAT-P-1b \citep{bnk+07,whb+07b}, WASP-1b
\citep{cww+07,cwe+07a}, TrES-4b \citep{moc+07}, XO-3b
\citep{wht+08}, CoRoT-Exo-2b \citep{aab+08}.  Giant transiting
planets can thus be highly sensitive diagnostics of the validity or
otherwise of specific theoretical predictions, but only if their
masses and radii can be measured with accuracies of 1--2\%. As new
detections of smaller transiting planets are expected over the next
few years, maintaining this level of accuracy will become more and
more challenging. \citet{skh07} used simple solid planet structure
calculations to show that uncertainties of a few \% at most are needed
to distinguish between different bulk compositions for sub-Uranus
planets using their location in the mass-radius plane -- whilst more
detailed calculations show that the mass-radius relation of planets
with mixed compositions may be degenerate \citep{ase08}.

A rapid review of the factors contributing to these uncertainties is
helpful at this stage. High precision planetary transit observations
allow the direct measurement of the planet-to-star radius ratio
$R_{\rm p}/R_{\rm s}$, the system scale (ratio of the semi-major axis
to the stellar radius $a/R_{\rm s}$), and the impact parameter $b$ (
$\equiv a/R_{\rm s} \cos i$, where $i$ is the inclination of the
orbit).

On the other hand, radial velocity observations of the host star yield
a measurement of the planet mass $M_{\rm p}$ relative to the star mass
$M_{\rm s}$, convolved with the inclination term $\sin i$ which is
known in the case of transits. For a circular orbit:
\begin{equation}
\frac{M_{\rm p}}{M_{\rm s}^{2/3}} = \frac{K}{\sin i}
 \left(\frac{P}{2\pi G}\right)^{1/3}
\label{eq:mp}
\end{equation}
where $K$ is the semi-amplitude of the host star's radial velocity
variations, $P$ is the orbital period of the system, $G$ is Newton's
constant of gravitation, and $M_{\rm p} \ll M_{\rm s}$. Combined fits
to the transit and radial velocity data can thus be used to measure
the relative planet mass and radius ratios.

Obtaining absolute estimates for these parameters requires knowledge
of the star mass and radius, which are usually obtained by comparing
high-resolution, high signal-to-noise ratio spectra of the star to
theoretical models. The transit observations provide a constraint on
$M_{\rm s}^{1/3}/R_{\rm s}$, which can be combined with the spectral
parameters of the star and theoretical evolutionary tracks to obtain
estimates of $M_{\rm s}$ and $R_{\rm s}$.

There are thus three classes of sources of uncertainty on the masses
and radii of transiting planets: those arising from the transit fit,
those arising from the radial velocity fit, and those arising from the
spectral analysis of the host star. The detailed listing of the
parameters of published transiting planets and their host stars,
maintained at {\tt http://www.inscience.ch/transits/} by F.\ Pont, can
be used to perform a basic evaluation of their relative impacts.  The
combined uncertainties on published transiting planet parameters vary
widely, from $\sim 1$\% to 11\% for $R_{\rm p}$ and to 20\% for ${M_{\rm
    p}}$, depending primarily on the brightness of the host star and
the degree to which each system has received detailed characterisation
using dedicated follow-up observations. Uncertainties on $M_{\rm s}$
and $R_{\rm s}$, which can reach up to 13\% and 17\% respectively in
some cases, are usually the dominant source of uncertainty for massive
(Jupiter-mass and above) planets, which represent the vast majority of
the transiting planet crop to date. However, as smaller and lower-mass
planets become increasingly detectable thanks to space-based transit
searches and improvements in ground-based radial velocity instruments,
the uncertainties arising from the transit and radial velocity fits
are expected to become more important. 

A specific problem arises when the transits become comparable in depth
with the amplitude of the intrinsic brightness fluctuations of the
host star. These are due to the temporal evolution and rotational
modulation of structures on their surface (stellar spots, plages,
granulation). The amplitude of these variations can be several orders
of magnitude greater than the transit signal, particularly for
terrestrial planets and/or active stars, and they can occur on
timescales significantly shorter than the orbital period of the planet
(see Fig.~\ref{fig:LC}, upper curve). Stellar variability can thus
hinder the detection of planetary transits \citep{afg04}, but a number
of `pre-detection' filters have been developed to tackle this
problem. The performance of several of these filters in terms of
transit detection was evaluated in the context of first CoRoT blind
test (\citealt{mpb+05}, hereafter M05), a hare-and-hounds exercise
involving 1000 simulated CoRoT light curves containing various
transit-like signals, stellar variability and instrumental noise. This
test showed that the most successful filters recover a detection
threshold close to that obtained in the presence of instrumental noise
only, except for a few cases involving the most active and rapidly
rotating stars simulated.

However, these filters also have the property of modifying the shape
of the transit signal (M05, \citealt{bl08} hereafter BL08), and would
destroy any signal at the period of the transit occuring on longer
timescales than a few hours. We therefore set out to develop an
alternative algorithm, hereafter referred to as `reconstruction
filter', designed to remove variability at other periods than that of
the transit but preserve the transit signal, once that period has been
determined. We then evaluated the resulting improvement in planet
parameter measurements compared to a benchmark pre-detection filter.

After briefly introducing the simulated dataset used for test purposes
throughout the paper in Sect.~\ref{sec:data}, we describe some of the
current pre-detection filters in Sect.~\ref{sec:curmeth}, and quantify
the effect of a benchmark filter on the transit signal. We then
describe the reconstruction filter and evaluate its effect on the
transit signal in the same way in Sect.~\ref{sec:itfilt}. The impact
of the two types of filter on the accuracy of planet parameter
measurements are compared in Sect~\ref{sec:plparam}, and the main results
are summarised in Sect~\ref{sec:conc}.

\section{The CoRoT blind test dataset}\label{sec:data}

\subsection{The original dataset}

\begin{figure*}
\centering
\includegraphics[width=\linewidth]{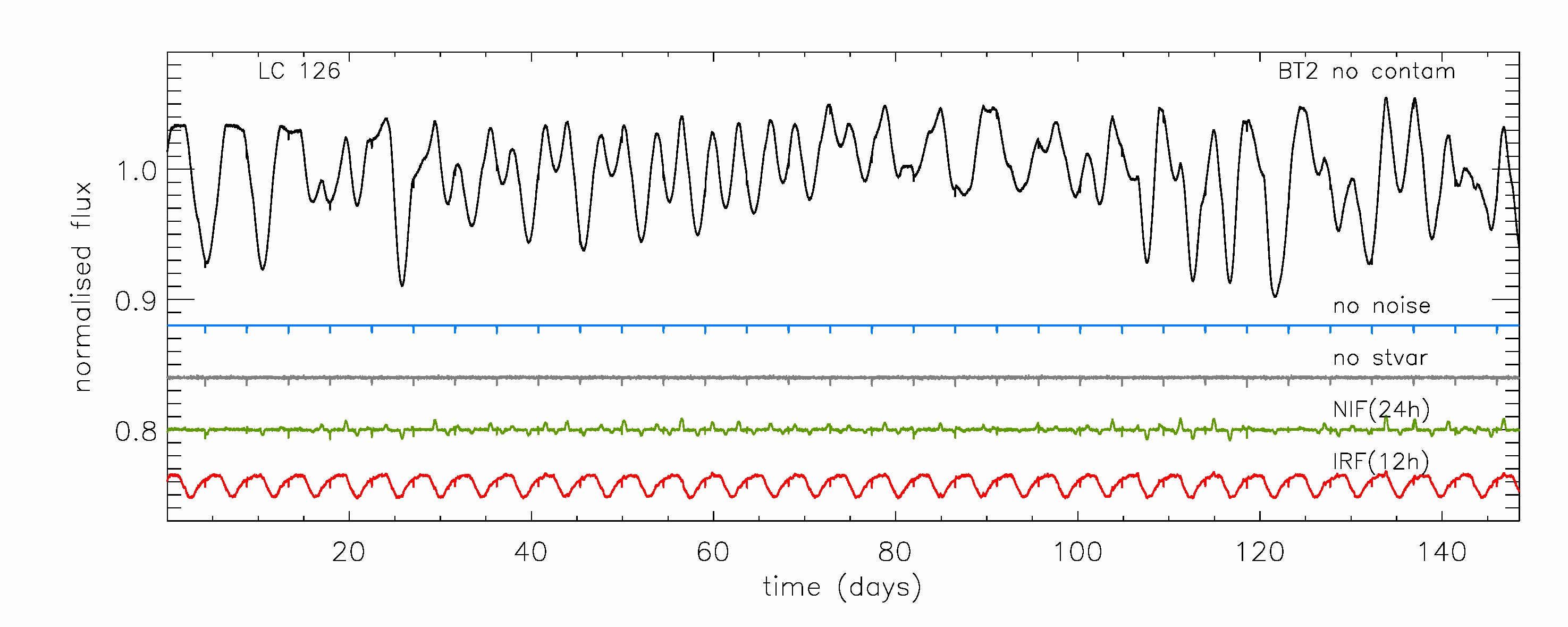}
\caption{Example of a simulated CoRoT BT2 light curve, corrected from the contaminant flux (2\%) coming from a background star (top curve,
  black line). This light curve contains the transit signal of a
  Saturn-like planet orbiting a particularly active Sun-like host star
  with an orbital period of 4.576\,d. Below the BT2 light curve are
  shown, from top to bottom and with a vertical offset for clarity,
  the no-noise (transits only, blue), no stellar variability (transits
  plus instrumental noise only, grey) light curves, as well as the
  NIF-filtered (green) and IRF-filtered (red) versions of the same
  light curve (see text for details).}
 \label{fig:LC}
\end{figure*}

The starting dataset used in this study is a sample of 26 simulated
CoRoT light curves with planetary transits taken from the second CoRoT
blind test (hereafter BT2; \citealt{maa+07}), which was carried out to
compare methods for discriminating between planetary transits and
grazing or diluted stellar eclipses. The production of the light
curves followed roughly the same steps as that for the first CoRoT
blind test (BT1), described in detail in M05, incorporating transits
simulated with the Universal Transit Modeler (UTM)\footnote{See {\tt
    http://www.iac.es/galeria/hdeeg/}.}, instrumental noise simulated
using the CoRoT instrument model \citep{abb+03}, and stellar
variability curve simulated using a combination of the methods of
\citet{lrp04} and \citet{afg04}.

An updated version of the CoRoT instrument model was used in the BT2,
incorporating more realistic satellite jitter and enabling the
production of 3-colour light curves, though the 3 bandpasses were
summed in the present study to construct a `white' light curve. The
two approaches used in the BT1 to model stellar variability were
merged in the BT2, using the scaled spot model of \citet{lrp04} to
simulate rotational modulation of active regions and the stochastic
model of \citet{afg04} to simulate granulation. The simulated transits
correspond to planet radii ranging from 0.2 to $1.1\,R_{\rm Jup}$,
orbital periods from 2.6 to 11.0\,d, and impact parameters from 0.25
to 0.88. As in the BT1, the flux in each aperture was modelled as
arising from two stars, only one of which contained a transit-like
signal. This is to reflect the fact that there is almost always one or
more backgrounds star in the CoRoT aperture. This has the effect of
diluting the transit signal, and to account for it we subtracted from
each BT2 light curve a constant corresponding to the fraction of the
median flux contributed by the star which is not eclipsed (see Tab.~\ref{tab:contam} in Appendix for contaminant fluxes (percentages of total flux) corrected from in each BT2 light curve studied). An example
of a light curve with transit from the BT2 is shown in
Fig.~\ref{fig:LC}, and a phase-folded version in
Fig.~\ref{fig:LCnc}. The full set of light curves is shown in
Appendix~\ref{app:lcs} (Figs.~\ref{fig:lcs1} \& \ref{fig:lcsad})

\subsection{The reference sets}\label{sec:refset}
\begin{figure*}
\centering
\includegraphics[width=\linewidth]{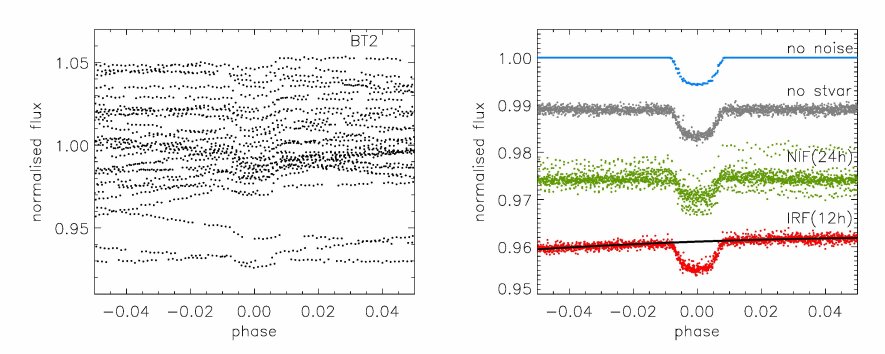}
\caption{Phase-folded plots of the light curves shown in
  Fig. \protect\ref{fig:LC}. Left: original BT2 light curve. Right:
  from top to bottom, no noise (blue), no variability (grey),
  NIF-filtered (green) and IRF-filtered (red). The black line overplotted on the IRF-filtered light curve is a $2^{nd}$ degree polynomial fit about the transit, by which the filtered light curve is divided before fitting the transit.}
 \label{fig:LCnc}
\end{figure*}

As we are using simulated data, each component of the signal is known
and can be studied individually. We have thus constructed two sets of
reference light curves, using on one hand only the transit signal (no
noise, no stellar variability) and on the other hand the transit
signal and instrumental noise only (no variability). We use the first
set to evaluate the reference values of the parameters derived from
transit fits in Sects.~\ref{sec:curmeth} to \ref{sec:plparam}. These
could have simply been deduced from the input parameters given to the
transit modelling software UTM when simulating the light
curve. However, there can be differences between those and the
parameters recovered from the transit fit due to the fitting process,
rather than to the noise, and we wish to keep those effects, which are
not specifically of interest here, separate from the effects of the
stellar and instrumental noise. The second set was used to provide a
benchmark for how well one can measure the parameters of interest in
the presence of instrumental (white) noise, i.e.\ if the stellar
variability was removed perfectly. These reference sets are shown in
blue and grey respectively in Figs.~\ref{fig:LC}, \ref{fig:LCnc},
\ref{fig:lcs1} and \ref{fig:lcsad}.

After visual analysis of our two reference sets of light curves, we
discarded two of the 26 light curves, where the transits were so small
as to be undetectable even in the light curves with no stellar
variability, as such cases would not realistically reach the
post-detection stage.

\section{Effect of pre-detection filters}\label{sec:curmeth}

\subsection{The benchmark pre-detection filter}\label{sec:NIF}

Pre-detection filters aim to remove stellar variability in light
curves to improve the detectability of transits, without any prior
knowledge of the transit signal except for the fact that stellar
variability typically occurs on longer time scales (hours to days)
than the transit signal (minutes to hours). All of the techniques
tested in the CoRoT BT1, which range from simple Fourier-domain
low-pass filters to slightly more sophisticated implementations
involving simultaneous fitting of hundreds of low-frequency sinusoids,
or time-domain nonlinear iterative filtering (\citealt{ai04},
hereafter AI04), exploit this difference. These filters proved
effective in removing stellar variability to facilitate the detection
of transits but, as pointed out in M05 and BL08, they deform the shape
of the transits. In this section, we quantify the impact of the
deformation cause by the nonlinear iterative filter (NIF) of AI04 on
the derived planet parameters. The NIF performance as a
pre-detection filter was recently compared to a range of other
published methods (BL08), and it emerged as the method of choice among
those compared. The NIF is used by the new filter described in this paper, to estimate the light curve continuum. These two reasons make the NIF a suitable benchmark for the present work, providing us with a direct evaluation of improvement in filtering performance.

The NIF separates stellar variability from the transit signal in the
time domain, using an iterative procedure with the following steps:
\begin{enumerate}
\item apply a short base-line (here we use 7 data-points, $\sim$1
  hour) moving median filter to smooth out the white noise and reduce
  the sharpness of any high-frequency features in the data;
\item apply a longer base-line (here we use 24 hours) moving median
  filter to the output of the step (i), followed by a shorter
  base-line (here we use 2 data-points, $\sim$17 minutes) boxcar
  (moving average) filter;
\item subtract the output of step (ii) from that of step (i) and
  evaluate the scatter of the residuals as $\sigma =1.48 \times {\rm
    MAD}$, where ${\rm MAD}$ is the median of the absolute values of
  the residuals;
\item flag all outliers differing by more than $n \sigma$ from the
  continuum;
\item return to step (ii) and repeat the process, interpolating over
  any flagged data points before estimating the continuum and
  excluding them when estimating the scatter of the residuals, until
  convergence is reached (typically less than 3 iterations);
\item subtract the final continuum from the original light curve.
\end{enumerate}
As the procedure converges, more and more of the in-transit points
become flagged at step (iv), so that the effect of the transits on the
final continuum estimate is minimal. However, the choice of long
base-line for the moving median filter in step (ii) and of $n$ in step
$(iv)$ must reflect a trade-off between appropriately following the
stellar variations and incorporating too much of the transit signal
when evaluating the continuum. This trade-off results in some of the
transit signal been unavoidably filtered along with the variability.
For the value of $n$ in step $(iv)$, one would normally use $n=3$ to
flag more in-transit points. In the case of the BT2, some light curves
contain very strong and rapid variability. Thus, using a low $n$ would
clip not only in-transit points but also out-of-transit points where
the variability is too rapid to be well modelled by the continuum
estimate (as in the example in Fig. \ref{fig:LC}). Hence, we used
$n=150$ in this paper, which effectively means no points are clipped
and convergence occurs at the first iteration.

\subsection{Quantitative impact on transit parameters}\label{sec:defNIF}

We applied the NIF (Section~\ref{sec:NIF}) to our sample of 24 BT2
light curves. The post-NIF light curves are shown in green on
Figs.~\ref{fig:LC}, \ref{fig:LCnc}, \ref{fig:lcs1} and
\ref{fig:lcsad}. Clear variability residuals are visible in the
unfolded post-NIF curves, corresponding to sections of the light curve
where the variability is too rapid to be filtered adequately. The
phase-folded light curves also show that the shape of the transits is
affected by the filter. In practical terms, the transit appears both
shorter and shallower than before filtering.

We then folded all light curves at the period of the injected transits
and performed least-squares fits of trapezoidal models to the results
to estimate the basic transit parameters: depth $\delta$, internal and
external duration $d_{\rm i}$ and $d_{\rm e}$ (respectively excluding
and including ingress and egress), and the phase $\phi$. The light
curves were normalised such that the out-of-eclipse level is always
1. The same folding and trapeze fitting procedure was applied to the
two reference sets described in Section \ref{sec:refset}. 

In 4 of the BT2 light curves (Fig. \ref{fig:lcsad}), the stellar variability was so strong that, after applying the NIF, the phase-folded transits were barely detectable, and meaningful fits to these transits impossible. These 4 light curves were excluded from the comparison of the NIF filter with noise- and variability-free cases which is discussed below.

\begin{figure}
\centering
\includegraphics[width=\linewidth]{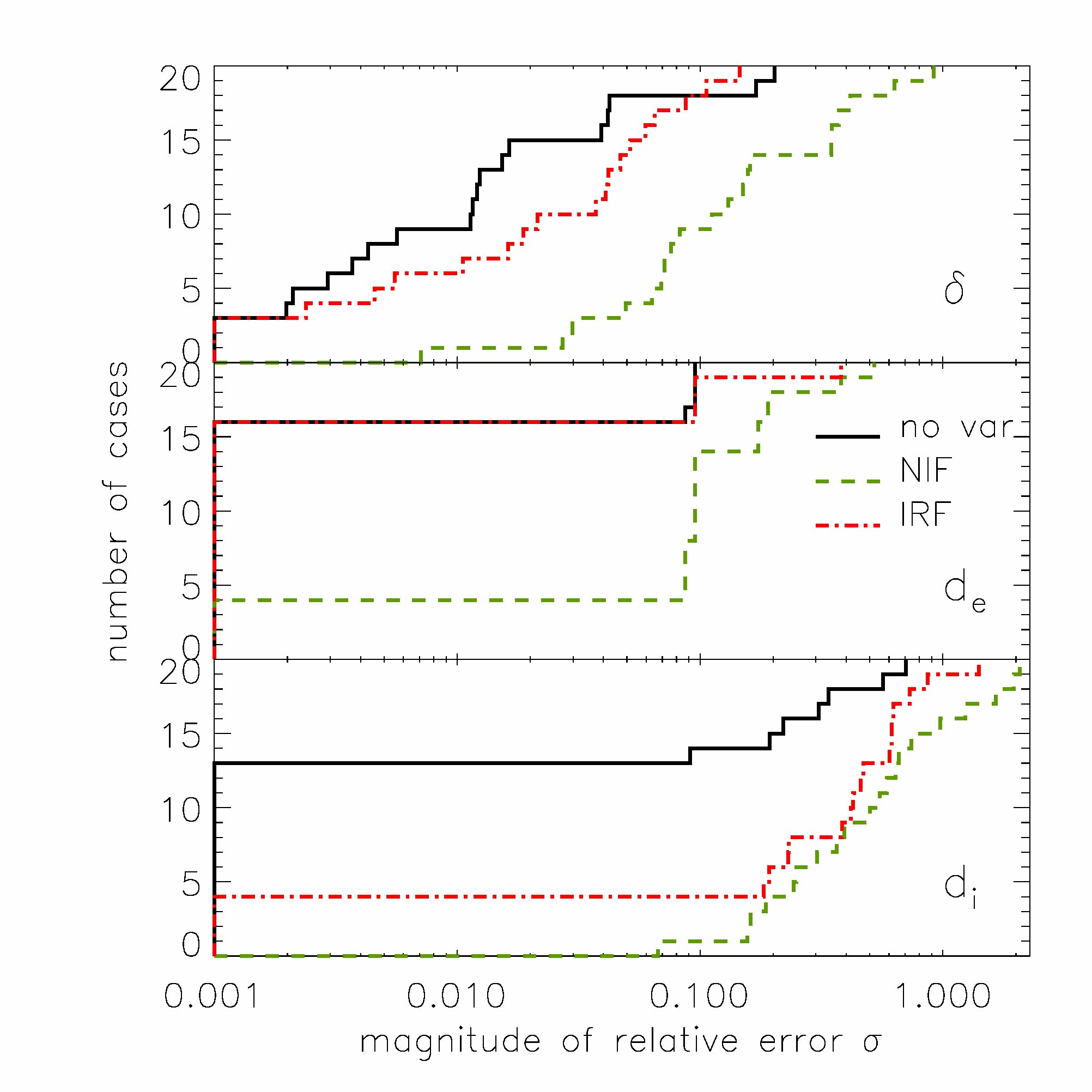}
\caption{Cumulative histograms of the relative error $\sigma$ (see
  text for exact definition) on the transit parameters measured from
  trapezoidal fits to the light curves with no variability (black),
  NIF-filtered light curves (green) and IRF-filtered light curves
  (red). Upper panel: transit depth $\delta$; middle panel: external
  duration $d_{\rm e}$ (total transit duration); lower panel: internal
  duration $d_{\rm i}$ (duration in full-transit).  $\sigma>1$ when a
  parameter is mis-estimated by more than its true value.}
 \label{fig:histdegrtr}
\end{figure}

We list the measured values of the transit parameters ($\delta$, $d_{\rm i}$ and $d_{\rm e}$) of direct relevance to the determination of planet parameters for all 20 light curves in Appendix~\ref{app:par} (Table~\ref{tab:trpara}). We also show, in Fig.~\ref{fig:histdegrtr}, cumulative histograms of the relative
error $\sigma(\theta) = |\theta - \theta_0| / \theta_0$, where
$\theta$ is the parameter of interest and the subscript $0$ refers to
the value measured from the reference light curve with transits only
(no noise), contrasting the NIF case (green dashed line) to the case
with no variability (black solid line). The median relative errors
obtained with the NIF over our sample are $\sigma_{\rm
  NIF}(\delta)=12\%$, $\sigma_{\rm NIF}(d_{\rm e})= 10\%$ and
$\sigma_{\rm NIF}(d_{\rm i})=52\%$, indicating that the planet
parameters would be seriously affected if derived from NIF-filtered
light curves. We note that the internal duration $d_{\rm i}$ tends to
be systematically underestimated even for the reference set of light
curves with no stellar variability. For a discussion of the source of
this well-known bias, see e.g.\ \citet{p08}.

\section{A post-detection reconstruction filter}\label{sec:itfilt}

\subsection{Definition}

In an attempt to avoid the undesirable effects of the NIF on the
transit shape, we developed an iterative reconstruction filter (IRF)
intended to remove stellar variability once the transits have been
detected, whilst altering the transit signal as little as possible.

The IRF is an iterative approximation of the full signal at the period of the transit. It uses the NIF to simultaneously estimate the continuum variation (i.e., stellar variability).

Let $\{Y(i)\}$ (where $i = 1$, \ldots, $N$, $N$ being
the number of data points in the light curve) represent the observed
light curve (which is assumed to be normalised), $\{A(i)\}$ the
detrended light curve and $\{F(i)\}$ the signal to be filtered out. We
give the main steps of the IRF below:
\begin{enumerate}
\item Select an initial estimate for $\{F(i)\}$. We adopt $\{F(i)\}=1$ as our initial estimate.
\item Compute a corrected time-series $\hat{Y}(i) \equiv Y(i) / F(i)$.
\item Estimate $\{\hat{A}(i)\}$ by folding $\{\hat{Y}(i)\}$ at the
  transit period and boxcar averaging it in intervals of a fixed duration in
  phase units (binning is used to reduce high frequency noise). For
  the BT2 light curves, a duration of 0.09\% of the phase was found to
  be suitable (this value was selected by trial and error, longer
  duration implying lower noise in the estimate of $\{\hat{A}(i)\}$
  but more distortion of the transit signal).
\item Unfold $\{\hat{A}(i)\}$ to obtain $\{{A}(i)\}$. Compute a new
  estimate of $\{F(i)\}$ by applying the NIF to $\{Y(i)/A(i)\}$. The baseline for the median filter
  used in the NIF at this step can be chosen on a case-by-case basis,
  and can be significantly shorter than in the pre-detection case,
  because it is applied to a light curve from which most of the
  transit signal has been removed. In the present study, we adopt a
  baseline of 12 hours.
\item Return to step (ii) with the new estimate of $\{F(i)\}$, and
  iterate until the condition $|\mathcal{D}_{j-1}-\mathcal{D}_{j}| <
  10^{-4}$ is satisfied for two consecutive iterations, where $j$ is
  the iteration number (initialisation at $j=0$), and
  \[ \mathcal{D}_{j} = \frac{\sum_{i=1}^{N} \left[ Y(i) / A_j(i) -
      F_j(i) \right]^2}{N-1}. \] 
In the case of the BT2 light curves, the convergence was reach after 4 iterations (i.e ${D}_{j}$ was calculated up to $j=6$).
\end{enumerate}
The final detrended light curve is given by $\{Y(i) / F(i)\}$, where
$\{F(i)\}$ is the last (presumably best) estimate of the stellar
variability.\\

\noindent This algorithm is in some ways analogous to the trend
  filtering algorithm (TFA) of \citet{kbn05} (hereafter KBN05) in
  post-detection mode. For clarity, we briefly list the main
  similarities and differences between the two algorithms. The TFA is
  designed to remove systematic trends common to large numbers of
  light curves in transit surveys, rather than stellar variability
  which is individual to each object, but both algorithms work by
  decomposing each light curve into three components: the signal of
  interest $\{A(i)\}$ (the transits), the signal to be filtered out
  $\{F(i)\}$ (the systematics in the case of the TFA and the stellar
  variability in the case of the IRF), and the residuals. In the TFA,
  the signal to filter out (systematics) is modeled as a linear
  combination of a number of template light curves selected from the
  survey sample. In the IRF, the signal to filter out (stellar
  variability) is taken as the continuum of the light curve estimated
  with the NIF (description in Section \ref{sec:NIF}). In this
  analogy, the NIF would be equivalent to TFA in pre-transit-detection
  mode. When used in reconstruction mode (post-detection), both
  methods make use of the knowledge of the transit period to
  iteratively improve the evaluation of the transit signal and of the
  signal to be filtered out (which is assumed not to be
  periodic). Whereas $\{F(i)\}$ and $\{A(i)\}$ are treated additively in the TFA,
  they are treated multiplicatively here since the signal to be
  filtered out is intrinsic to the star, and the planet hides a
  certain fraction of the flux emitted by the star. This also results
  in a different convergence criterion. In KBN05, the first estimate
  of $\{F(i)\}$ is obtained from the pre-detection implementation of the
  TFA. In the IRF, it would be counter-productive to use the
  NIF-filtered light curve as the initial estimate of $\{F(i)\}$,
  since we have shown that the NIF affects the transit signal we are
  trying to reconstruct (see Section \ref{sec:defNIF}), so the inital
  estimate of $\{F(i)\}$ is taken to be constant at 1. Finally, the IRF
  treats high frequency effects by smoothing the phase-folded signal,
  while the TFA treats them by filtering out common outlier values.

\subsection{Performance on simulated light curves}\label{sec:trpara}


The red curves in Figs.~\ref{fig:LC}, \ref{fig:LCnc}, 
\ref{fig:lcs1} and \ref{fig:lcsad}, show the light curves in our sample
after applying the IRF. Visually, one does not detect any sign of deformation of transit shape. On the other hand, a
different feature, which can be seen as a limitation of this filter,
is immediately apparent: the IRF preserves any signal at the period of the transit. This property has positive consequences: it implies that
potentially interesting signals, such as secondary eclipses, reflected
light variations, or thermal emission variations, are preserved. On
the other hand, any power in the stellar variability signal at the
frequency corresponding to the planet's orbital period is also
preserved. This could be reduced -- but not entirely eliminated -- by
taking the initial estimate of $\{F(i)\}$ to be the continuum
estimated with the NIF, but with a base-line long enough not to have a
significant impact on the transit signal (though it would remove any
slower phase variations associated with the planet). On the other
hand, it is straightforward to remove the residual variability about
the phase-folded transit, for instance by fitting a low-order
polynomial to the data about the transit in the final phase-folded
light curve. Dividing the final phase-folded light curve by this
polynomial allows to extract a normalised phase-folded transit that
can be used to derive the planet parameters.  Such a technique is
commonly applied to remove variability about each transit in
unfiltered, unfolded light curves (see e.g.\ \citealt{aap+08}). Using it
on the folded light curve after applying the IRF significantly reduces
the number of free parameters, and thus should increase the
reliability of the results. An example of such a fit is shown on
Fig.~\ref{fig:LCnc}.

The IRF was applied to our sample of 24 BT2 light curves, and the
residual variability around the transit was removed by subtracting a
$2^{nd}$ order polynomial fit of the continuum about the
transit. This re-normalisation is important as the trapezoidal
  fit function used assumes a constant out-of-transit level. The
transit parameters were then estimated from a trapezoidal fit to the
resulting phase-folded transit in the same way as described in
Section~\ref{sec:defNIF} for the NIF case. The results are listed in
Table~\ref{tab:trpara} and shown as the red dash-dot curves in
Fig.~\ref{fig:histdegrtr}. For the 20 BT2 transit light curves which
were also used to evaluate the performance of the NIF, the IRF gives
median relative errors of $\sigma_{\rm IRF}(\delta) = 3\%$,
$\sigma_{\rm IRF}(d_{\rm e})<10^{-4}\%$ and $\sigma_{\rm IRF}(d_{\rm
  i})=$ 42\%, representing a significant improvement over the NIF
case. Additionally, in the 4 cases where the transits were barely
detectable after applying the NIF, which are not included in the
comparison sample, the transits are clearly detectable and yield
meaningful fits after applying the IRF.

Looking at Fig.~\ref{fig:histdegrtr} in more detail, we see that,
where a relative error on the transit depth in excess of 10\%
(essentially precluding any meaningful constraints on the planet
structure) occurs in 60\% of the cases studied with the NIF, it occurs
in only 5\% of the cases with the IRF. Similarly, the NIF yielded
$\sigma(\delta)<3\%$ (potentially allowing discrimination between
different kinds of evolutionary models as well as a reliable basic
structure determination) in only 15\% of the cases, but the IRF did so
in 50\% of the cases. 

It is also clear that the external transit duration is recovered
near-optimally in the light curves treated with the IRF, with
$\sigma(d_{\rm e})<0.1\%$ in 80\% of the cases and $\sigma(d_{\rm
  e})<10\%$ in 95\% of the cases, compared to a significantly decreased performance with the NIF. However, although the IRF also
systematically improves the determination of the internal transit
duration compared to the NIF, this improvement is much less
significant, and the relative errors remain large (more than 10\% for 80\% of the cases studied). This implies that the IRF would probably not significantly increase the number of cases where both internal and
external duration can be determined precisely enough to break the
degeneracy between system scale and inclination, and thus to constrain
the stellar density in a model-independent fashion.

\section{Implications for star-planet parameters}\label{sec:plparam}

\begin{figure}
\centering
\includegraphics[width=\linewidth]{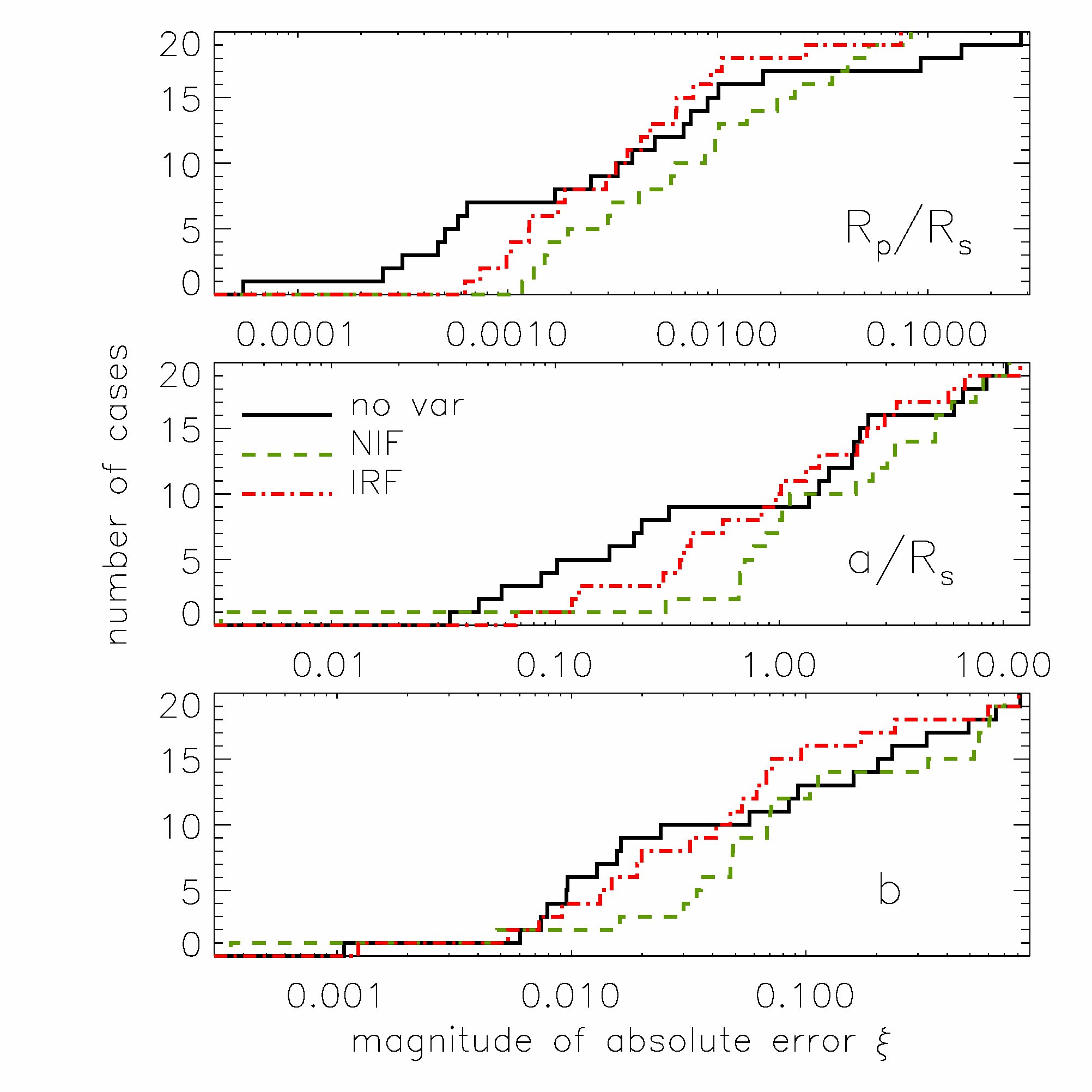}
\caption{Cumulative histograms of the errors $\xi$ (see text for
  definition) for star-planet parameters measured from MA02 transit
  fits to the light curves with no variability (black), NIF-filtered
  light curves (green) and IRF-filtered light curves (red). Upper
  panel: radius ratio $R_{\rm p}/R_{\rm s}$;
  middle panel: system scale $a/R_{\rm s}$;
  lower panel: impact parameter $b$.}
\label{fig:histdegrpl}
\end{figure}

Although the basic trapezoidal fits performed in the previous two
sections provide a quick estimate of the degree of deformation of the
transit signal due to the variability filtering process, one would in
practice perform a full transit fit based on a physical model of the
star-planet system. \citet{ma02} (hereafter MA02) provided an
analytical formulation which has become very widely used for such
purposes, and was also used to generate the transits injected in the
BT2 light curves. 

By definition, the IRF preserves any signal at the period of the
  transit. If the stellar variability contains power at this period,
  it is preserved, inducing a flux gradient around the transit. This
  must be removed before fitting, since the MA02 formalism assumes
  that the out-of-transit level is 1. This was done by fitting a
  $2^{\rm nd}$ order polynomial fit -- the lowest-order found to give
  satisfactory results -- to the data about the transit in the
  phase-folded curve (based on two segments, each lasting 0.1 in
  phase, and offset by 0.15 in phase from the center of the transit on
  either side) before fitting the transits. Note that this is still a
  significant improvement over the common practice of performing a
  local polynomial fit to the vicinity of each transit, since the
  latter option has many more free parameters (one set of free
  polynomial parameters per transit, rather than one for the entire
  light curve).

We then used the quadratic limb darkening prescription of MA02 to
fit transit models to the 20 BT2 transit light curves where the
transits were clearly detectable with both filters. We performed these
fits on both reference sets described in Section~\ref{sec:refset} (no
noise and no variability), as well as on the BT2 light curves
themselves after applying the NIF on the one hand, and the IRF
followed by a polynomial fit to the region around the transit on the
other hand. The fits were performed using an {\sc idl} implementation
of the Levenberg-Marquart algorithm. The parameters of the model used
are the transit epoch $T_0$, the period $P$, the system scale
$a/R_{\rm s}$ (where $a$ is the semi-major axis), the star-to-planet
radius ratio $R_{\rm p}/R_{\rm s}$, the orbital inclination $i$ (or
impact parameter $b \equiv a \cos i / R_{\rm s}$), and the quadratic
limb-darkening coefficients $u_a$ and $u_b$. In this study, we fixed
the period and limb-darkening coefficients at the values used to build
the light curves\footnote{Visual examination of the phase-folded
  light curves revealed that the folding was not perfect even in the
  no noise case, suggesting that the period values used may have been
  slightly inaccurate. We attempted to refine the periods but did not
  succeed. It seems that the observation dates in the light curve
  files themselves, rather than the periods, suffer from a small
  rounding error. It is not possible to remedy this problem without
  re-generating the entire light curve set, but it is not expected to
  affect the results strongly, and any effect would be common to all
  versions of a given light curve.}. The initial epoch was taken
directly from the trapezoidal fits. The initial value for $a/R_{\rm
  s}$ was derived from the period using Kepler's $3^{\rm rd}$ law,
assuming $R_{\rm s}=R_{\odot}$ and $M_{\rm s}=M_{\odot}$. In order to
ensure convergence in both grazing and full transits we selected,
after some trial and error, an initial inclination of 83.7$^{\circ}$. We assumed zero eccentricity in all cases
(all the transit light curves in our sample were simulated for
circular orbits).

The results of the fits are listed in Table~\ref{tab:plpara}, while
the fits themselves are shown in Figs.~\ref{fig:lcs1} and \ref{fig:lcsad}. They are also compared in cumulative histogram form in Fig.~\ref{fig:histdegrpl} (with the same colour and line scheme as Fig.~\ref{fig:histdegrtr}). Instead of the relative error $\sigma$, we show the absolute error $\xi=|\theta-\theta_0|\equiv\sigma \times \theta_0$ with respect to the no noise case (subscript 0), for $\theta$ the key planet parameters $R_{\rm p}/R_{\rm s}$, $a/R_{\rm s}$ and $b$.

The IRF provides an overall improvement over the NIF in all three
parameters, reducing the median of $\xi(R_{\rm p}/R_{\rm s})$ from
0.007 to 0.003, $\xi(a/R_{\rm s})$ from 1.7 to 1.0, and $\xi(b)$ from
0.07 to 0.04 for $b$. For comparison, the corresponding median values
for the the case with no variability are 0.003, 1.4 and 0.07
respectively. However, the situation is not as clear cut as when
viewed in terms of transit parameters: there are a few cases where the
NIF gives a better match with the parameters obtained from the
noise-free light curves, and even cases where the largest error occurs
in the light curves containing instrumental noise only. 

In an attempt
to understand the reason for this, we examined all the light curves
one by one (see Figs.~\ref{fig:lcs1} and \ref{fig:lcsad}). The light
curves separate fairly naturally into three broad classes:
\begin{enumerate}
\item cases where the IRF performed better than the NIF (transit shape
  and derived planet parameters closer to the shape and parameter
  obtained in the absence of stellar variability): light curves 126,
  162, 169, 196, 200, and 223. These are cases where the original  
  light curves contain large amplitude, short timescale stellar
  variability (active and rapidly rotating stars).
\item cases where the NIF performance was already satisfactory, and
  the IRF gives results similar to the NIF: light curves 145, 152,
  186, 193, 208, 225, and 233.
\item cases where, while the transit reconstructed with the IRF
  appears closer to the original than the transit in the NIF-treated
  curve, the fitted parameters are not significantly improved or even
  worse: light curves 131, 133, 135, 154, 177, 192, 220. These are
  typically low signal-to-instrumental noise transits, where it
  becomes difficult to break the degeneracy between impact parameter
  and system scale. The radius ratio is typically less affected,
  except in the highest impact parameter cases (grazing transits).
\end{enumerate}
Thus, we can see that where the limiting factor was stellar
variability, the IRF is very successful in improving the errors on the
planet parameters. As might be expected, the improvement is minor or
non-existent where the limiting factor was the signal-to-white noise
or the grazing nature of the transits.

\section{Conclusions and future work}\label{sec:conc}

The transit and the stellar signal cannot be separated effectively if
they overlap too much in the frequency domain. Because of this,
commonly used pre-detections stellar variability filters, such as the
NIF, alter the transit signal, causing systematic errors in the
resulting star and planet parameters. We have quantified this effect
using 20 CoRoT BT2 simulated light curves including transits,
instrumental noise and stellar variability. We found that the effect
on the transit signal can be very significant, leading to errors on
the star-planet radius ratio up to 50\%.

We thus developed the IRF to take advantage of the strictly periodic
nature of planetary transits (in the absence of additional bodies in
the system) to isolate the transit signal more effectively, following
a method similar to the TFA algorithm previously developed for the
reconstruction of transits in the presence of systematics. We
evaluated the performance of the IRF relative to the NIF and the no
variability light curves by comparing a) the transit parameters from
trapezoidal fits, b) the star-planet parameters from analytic transit
fits, and c) the light curves themselves by visual examination. The
results can be summarised as follows: the transits reconstructed with
the IRF are systematically closer to the no variability case than the
NIF-processed transits, and the improvement in the transit depth and
duration can be very significant particularly in cases with large
amplitude, high frequency stellar variability. However, the full
transit fits are affected by other factors including instrumental
noise and the well known degeneracy between system scale and impact
parameter, which dominate the final parameter estimates in
approximately one third of the cases in our sample, or about half of
the cases where the IRF provided a visual improvement over the NIF.

In the near future, we intend to test the IRF on the light curves of
confirmed planets detected by the CoRoT mission, particularly those
orbiting active stars, in an attempt to refine the planet parameters,
but also to search for reflected light, primarily in the form of
secondary eclipses. As for the primary transit, polynomial fits about
the putative secondary eclipse location can be used to remove residual
stellar variations at the period of the transit.

Another potential application of the IRF occurs at the detection
stage. Among the 24 light curves of our sample, we already mentioned
that there were 4 where the residual stellar variability after NIF
filtering was too strong to perform any kind of meaningful
fit. Naturally, these events were not detected during the original
blind test for which the light curves were generated. There are two
more cases which we did include in our 20-strong comparison sample, as their transits after NIF-filtering could still be fitted,
but which transits were also not detected in the original exercise: light curves 192 and 200. After applying the IRF, two of these 6 cases became detectable\footnote{The detectability of the events was
  evaluated using the transit search algorithm of AI04, which was used
  in both CoRoT blind tests.} (light curves 165 and 200), the other 4 cases stayed undetectable due to the level of instrumental noise. Using the IRF as part of the detection process might therefore enable the detection of transits which would otherwise be missed around
particularly active stars. However, since the IRF would have to be run
at each trial period, and is relatively computationally intensive,
this would require a very large amount of CPU time unless the
algorithm can be significantly optimised. As radial velocity
measurements are also affected by stellar activity (which induces
radial velocity jitter and line bisector variations at the rotation
period of the star), it is not clear at this stage that the CPU
investment needed would be justified.

\section*{Acknowledgements}

The authors are grateful to Tim Naylor and Frederic Pont for useful
discussions and comments, and to Claire Moutou for providing the BT2
source material and patiently answering many queries. This work made
use of the {\sc mpfit} software written and made available online by
C.\ Markwart. We would also like to thank the referee for their
careful reading of the manuscript and constructive comments.

\bibliographystyle{mn2e} \bibliography{IRF}

\appendix

\section{Full light curve sample}
\label{app:lcs}

\begin{figure*}
\centering
\includegraphics[width=0.97\linewidth]{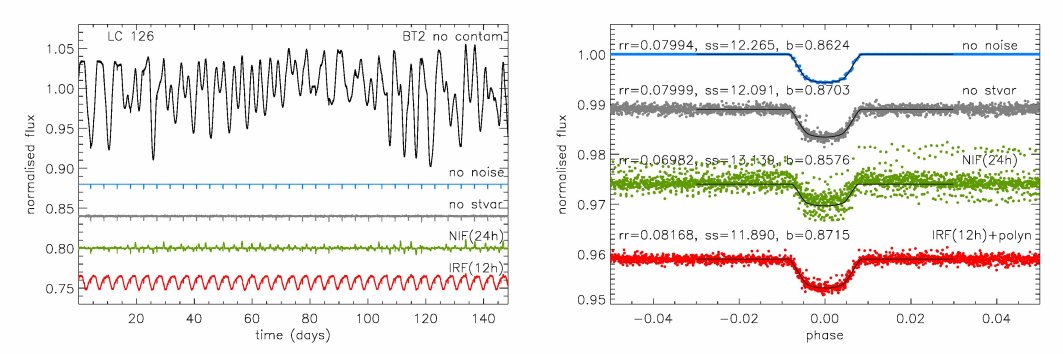}
\includegraphics[width=0.97\linewidth]{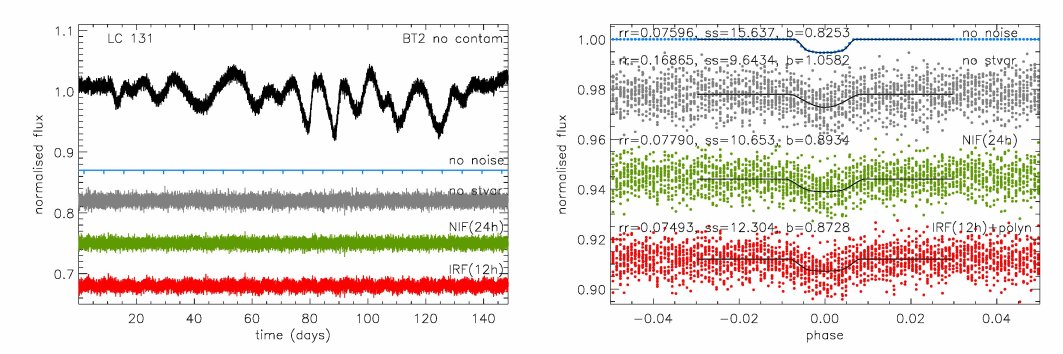}
\includegraphics[width=0.97\linewidth]{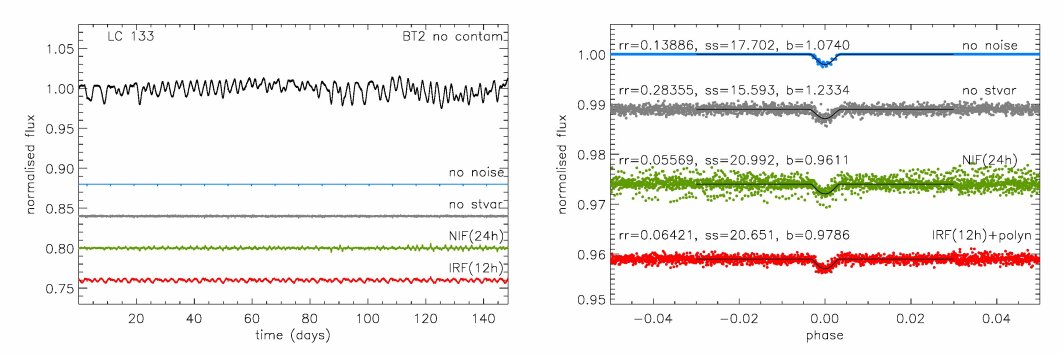}
\includegraphics[width=0.97\linewidth]{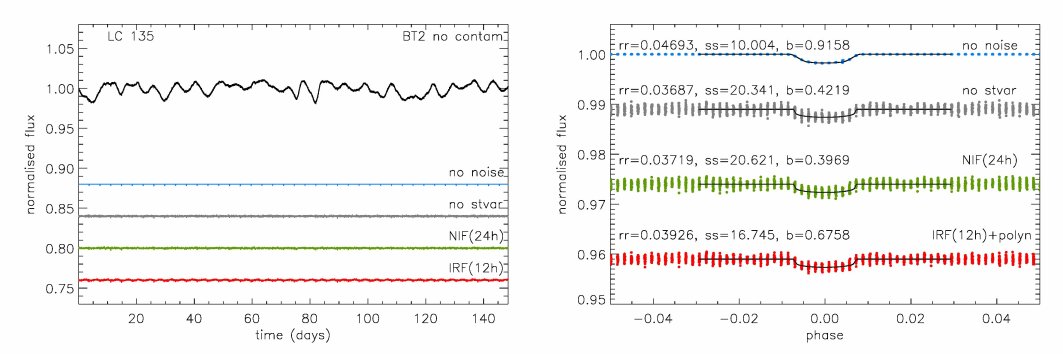}
\caption{The 20 BT2 light curves in the comparison sample. The light
  curve number is shown on the plots in the left column (original BT2
  numbering scheme) and the planet to star radius ratio ({\tt rr}),
  system scale ({\tt ss}), and impact parameter ({\tt b}) in the right
  column.}
\label{fig:lcs1}
\end{figure*}

\begin{figure*}
\centering
\includegraphics[width=0.97\linewidth]{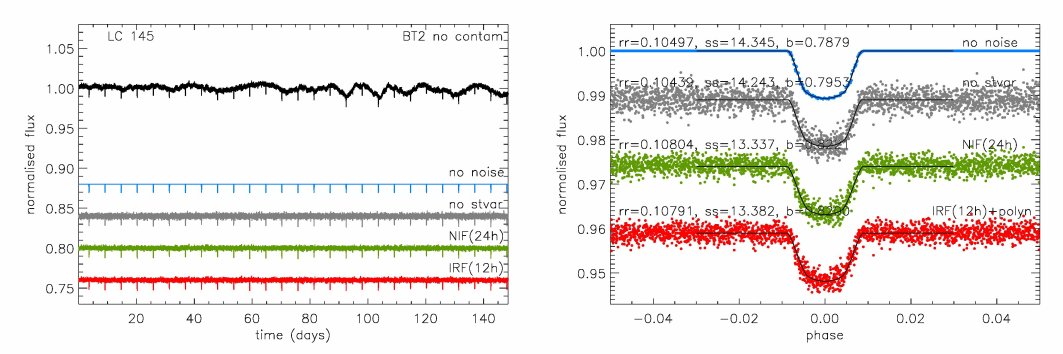}
\includegraphics[width=0.97\linewidth]{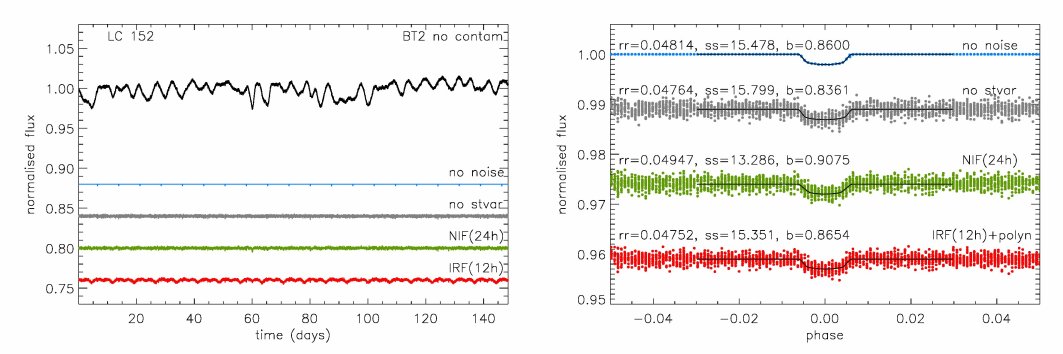}
\includegraphics[width=0.97\linewidth]{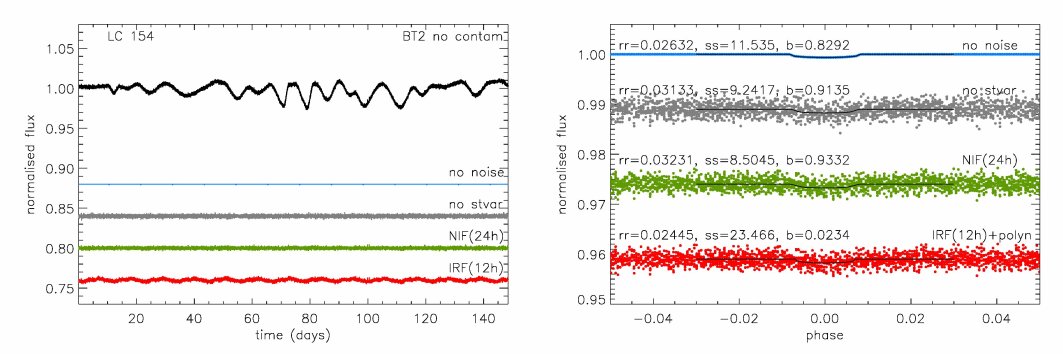}
\includegraphics[width=0.97\linewidth]{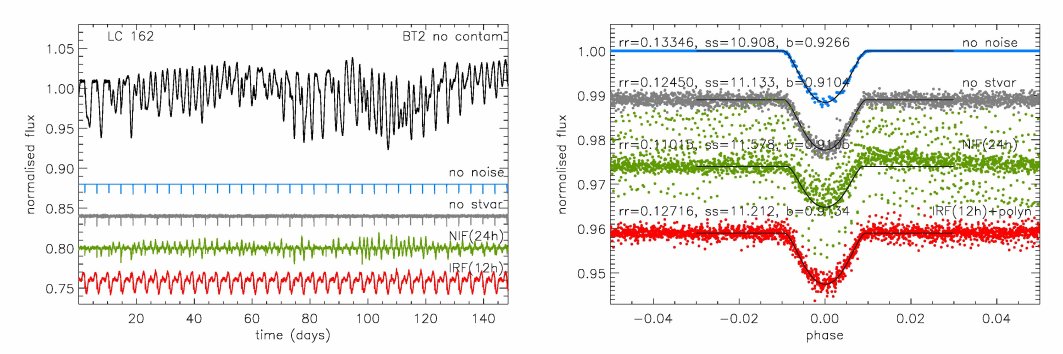}
\contcaption{}
\end{figure*}

\begin{figure*}
\centering
\includegraphics[width=0.97\linewidth]{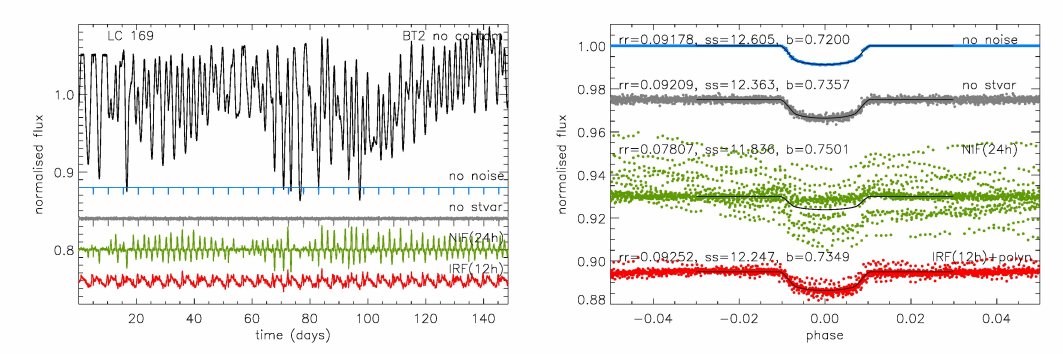}
\includegraphics[width=0.97\linewidth]{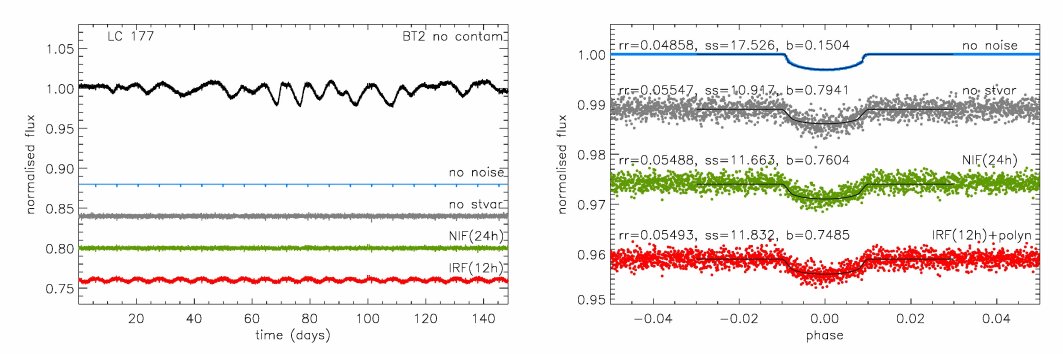}
\includegraphics[width=0.97\linewidth]{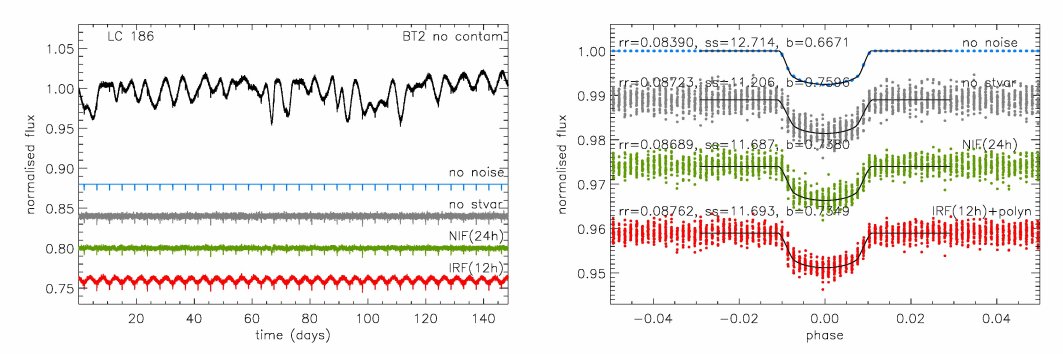}
\includegraphics[width=0.97\linewidth]{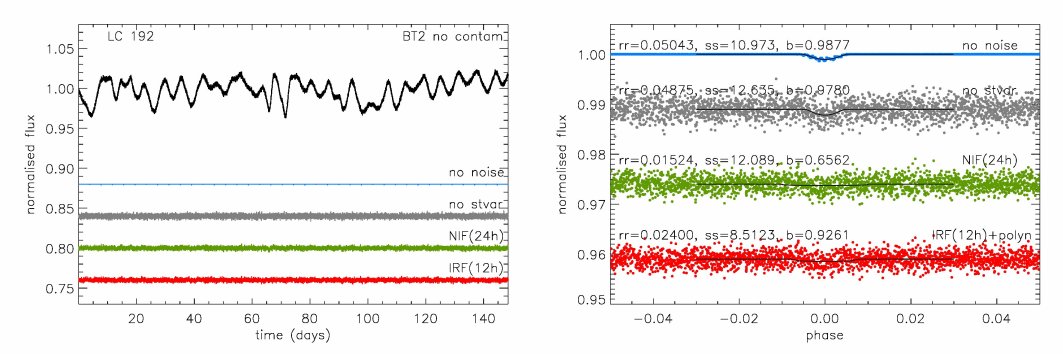}
\contcaption{}
\end{figure*}

\begin{figure*}
\centering
\includegraphics[width=0.97\linewidth]{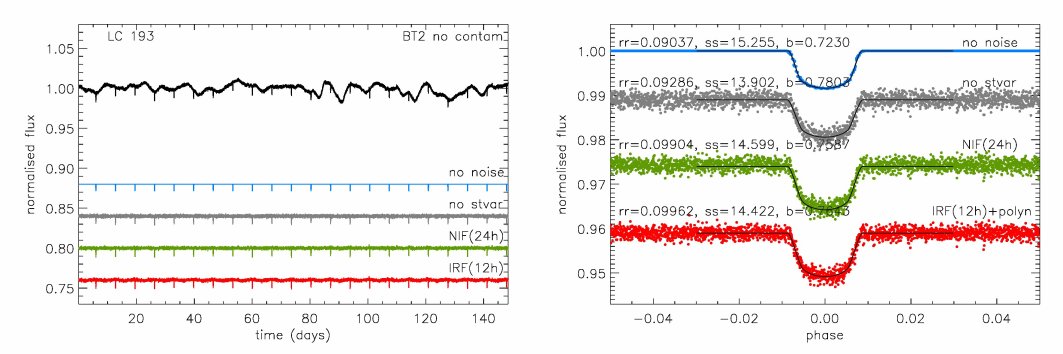}
\includegraphics[width=0.97\linewidth]{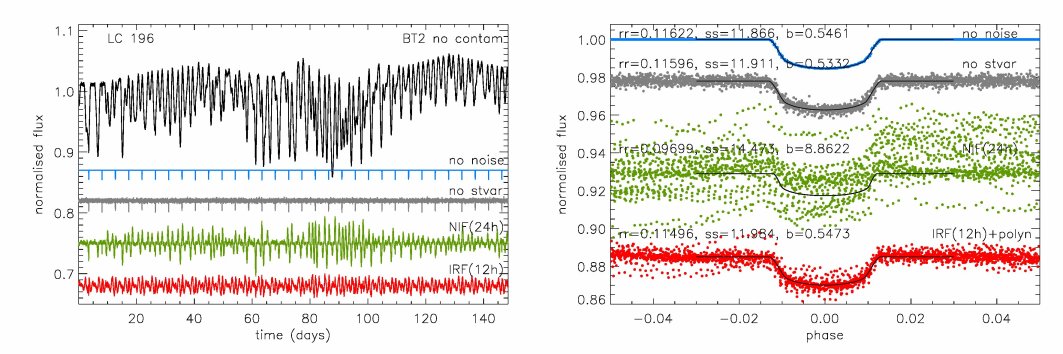}
\includegraphics[width=0.97\linewidth]{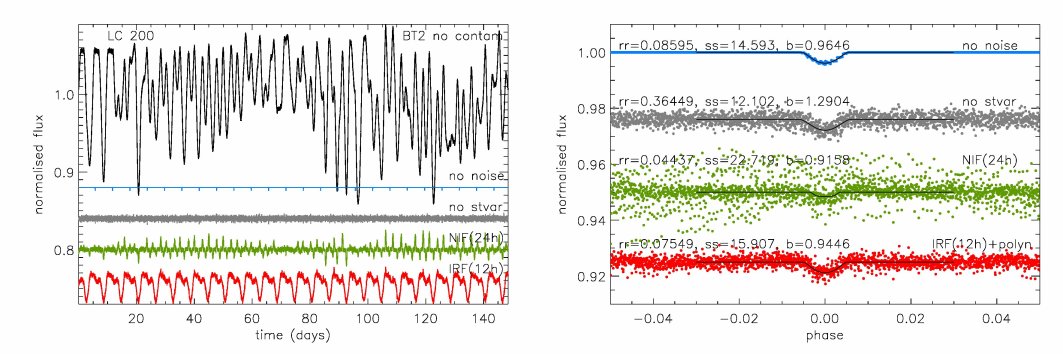}
\includegraphics[width=0.97\linewidth]{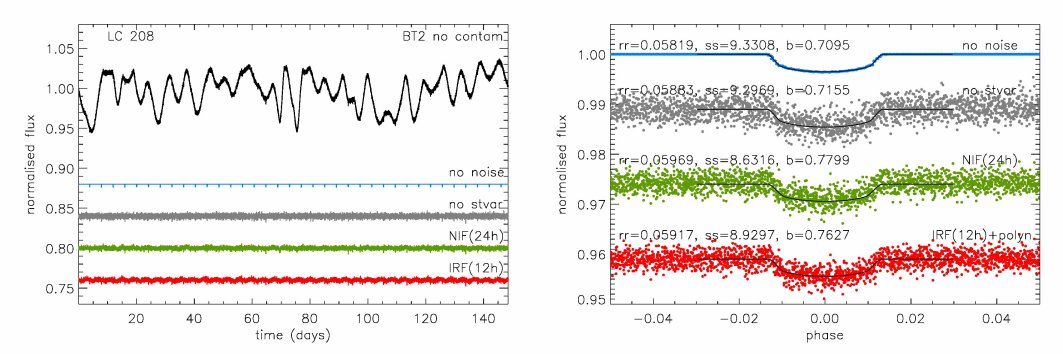}
\contcaption{}
\end{figure*}

\begin{figure*}
\centering
\includegraphics[width=0.97\linewidth]{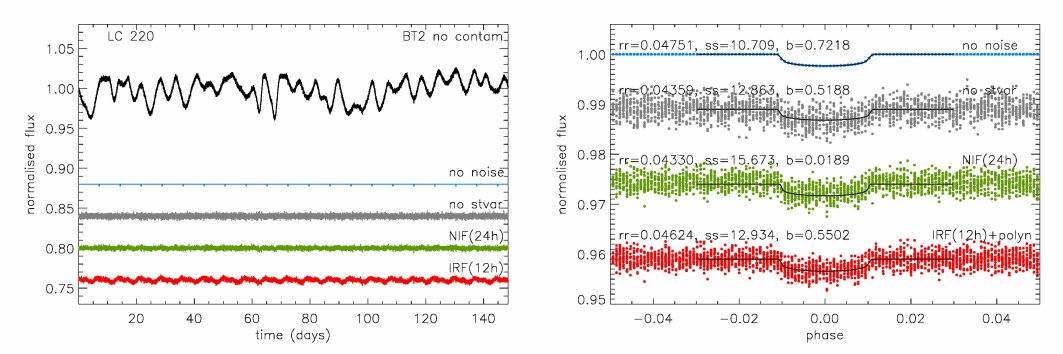}
\includegraphics[width=0.97\linewidth]{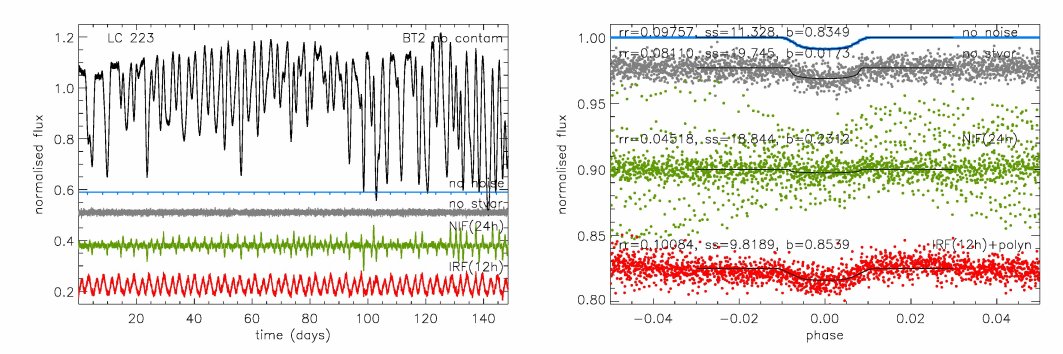}
\includegraphics[width=0.97\linewidth]{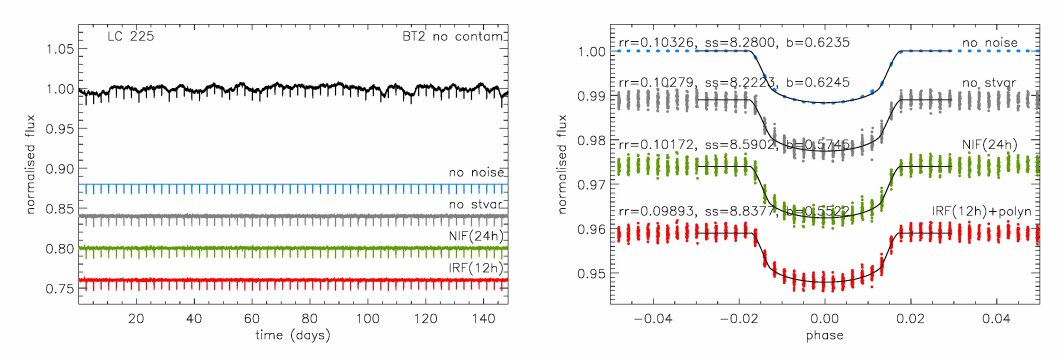}
\includegraphics[width=0.97\linewidth]{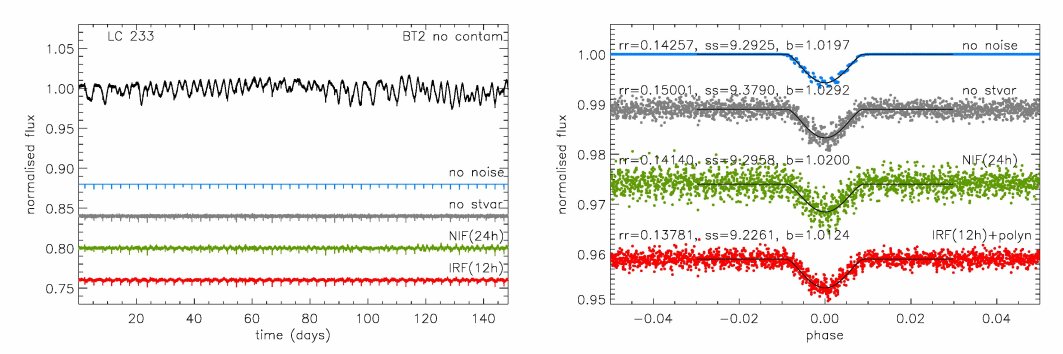}
\contcaption{}
\end{figure*}

\begin{figure*}
\centering
\includegraphics[width=0.97\linewidth]{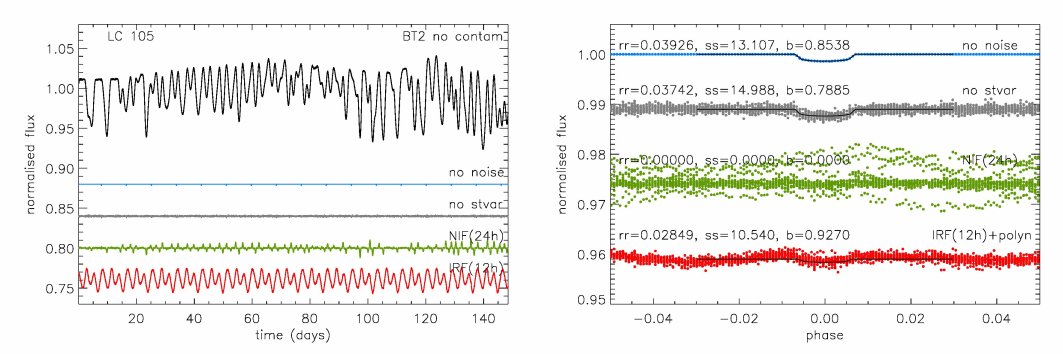}
\includegraphics[width=0.97\linewidth]{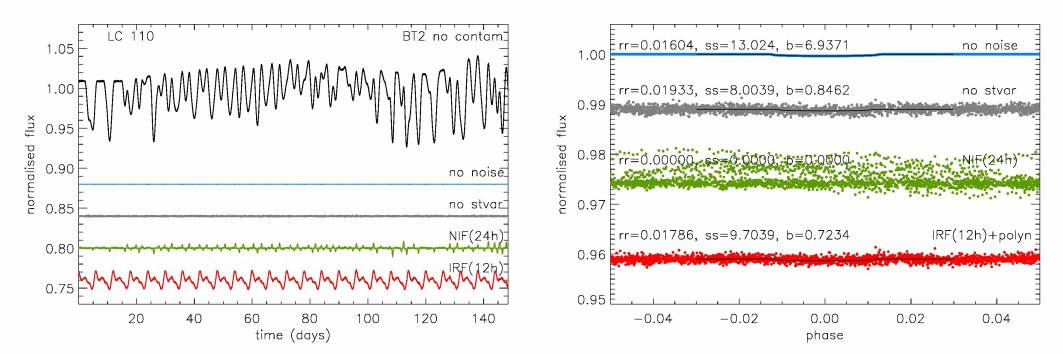}
\includegraphics[width=0.97\linewidth]{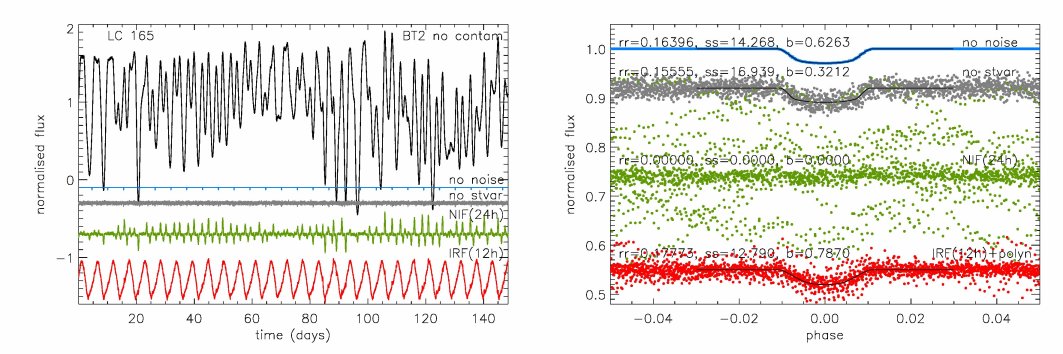}
\includegraphics[width=0.97\linewidth]{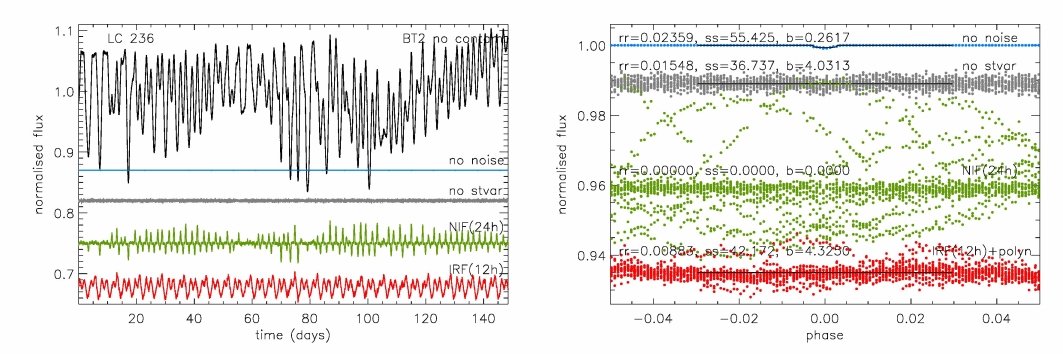}
\caption{The 4 BT2 transit light curves where the transit was
  undetectable after applying the NIF and miningful fits to the resulting transits were not possible. Same legend as
  Fig.~\ref{fig:lcs1}. These transits became boarder-line detectable in the IRF-filtered light curves.}
\label{fig:lcsad}
\end{figure*}

\section{Best-fit parameters}
\label{app:par}

\begin{table*}
  \caption{Transit parameters (transit depth $\delta$, total transit duration $d_{\rm e}$, and in-full transit duration $d_{\rm i}$) derived from trapezoidal fits to the light curves with transit signal only (`no noise'), transit signal and instrumental noise only (`no stvar'), the BT2 light curves filtered using  the pre-detection nonlinear iterative filter (`NIF'), and the same light curves filtered using post-detection iterative reconstruction filter (`IRF').}             
\centering          
\begin{tabular}{lrcccccccccccc}  
\hline\hline       
LC &  period & \multicolumn{4}{c}{$\delta$} & \multicolumn{4}{c}{$d_{\rm e}/P$} & \multicolumn{4}{c}{$d_{\rm i}$/P} \\
& (days) & no noise & no stvar  & NIF & IRF & no noise & no stvar  & NIF & IRF & no noise & no stvar & NIF & IRF \\
\hline 
\hline 
126 &  4.576  & 0.00501  & 0.00495  & 0.00326 & 0.00504 & 0.0153   & 0.0153   & 0.0138 & 0.0153 & 0.0064   & 0.0064   & 0.0084 & 0.0079 \\
131 &  6.880  & 0.00477  & 0.00469  & 0.00437 & 0.00448 & 0.0134   & 0.0121   & 0.0108 & 0.0121 & 0.0056   & 0.0017   & 0.0098 & 0.0098 \\
133 &  8.128  & 0.00168  & 0.00161  & 0.00155 & 0.00160 & 0.0058   & 0.0058   & 0.0047 & 0.0057 & 0.0016   & 0.0010   & 0.0026 & 0.0015 \\
135 &  3.733  & 0.00155  & 0.00148  & 0.00144 & 0.00152 & 0.0147   & 0.0147   & 0.0147 & 0.0147 & 0.0062   & 0.0062   & 0.0073 & 0.0090 \\
145 &  5.557  & 0.00938  & 0.00949  & 0.00931 & 0.00923 & 0.0167   & 0.0153   & 0.0167 & 0.0167 & 0.0054   & 0.0064   & 0.0067 & 0.0086 \\
152 &  7.360  & 0.00185  & 0.00185  & 0.00158 & 0.00176 & 0.0115   & 0.0125   & 0.0104 & 0.0115 & 0.0060   & 0.0065   & 0.0090 & 0.0071 \\
154 & 10.987  & 0.00056  & 0.00065  & 0.00054 & 0.00061 & 0.0172   & 0.0172   & 0.0155 & 0.0237 & 0.0088   & 0.0038   & 0.0067 & 0.0051 \\
162 &  4.171  & 0.00933  & 0.00922  & 0.00585 & 0.00894 & 0.0167   & 0.0167   & 0.0138 & 0.0167 & 0.0037   & 0.0037   & 0.0074 & 0.0070 \\
169 &  5.195  & 0.00772  & 0.00770  & 0.00504 & 0.00769 & 0.0209   & 0.0209   & 0.0191 & 0.0209 & 0.0066   & 0.0066   & 0.0109 & 0.0107 \\
177 &  7.339  & 0.00271  & 0.00267  & 0.00252 & 0.00260 & 0.0209   & 0.0209   & 0.0191 & 0.0209 & 0.0066   & 0.0046   & 0.0090 & 0.0107 \\
186 &  4.373  & 0.00683  & 0.00679  & 0.00649 & 0.00690 & 0.0209   & 0.0209   & 0.0191 & 0.0209 & 0.0087   & 0.0087   & 0.0134 & 0.0127 \\
192 &  3.915  & 0.00085  & 0.00102  & 0.00071 & 0.00076 & 0.0086   & 0.0094   & 0.0078 & 0.0078 & 0.0029   & 0.0023   & 0.0078 & 0.0070 \\
193 &  6.763  & 0.00749  & 0.00747  & 0.00847 & 0.00858 & 0.0167   & 0.0167   & 0.0167 & 0.0167 & 0.0070   & 0.0070   & 0.0065 & 0.0086 \\
196 &  4.608  & 0.01378  & 0.01384  & 0.00509 & 0.01288 & 0.0248   & 0.0248   & 0.0201 & 0.0248 & 0.0127   & 0.0127   & 0.0201 & 0.0175 \\
200 &  5.995  & 0.00317  & 0.00313  & 0.00185 & 0.00311 & 0.0095   & 0.0095   & 0.0059 & 0.0086 & 0.0023   & 0.0023   & 0.0052 & 0.0038 \\
208 &  4.064  & 0.00313  & 0.00301  & 0.00278 & 0.00302 & 0.0267   & 0.0267   & 0.0242 & 0.0267 & 0.0136   & 0.0136   & 0.0158 & 0.0162 \\
220 &  7.253  & 0.00215  & 0.00212  & 0.00181 & 0.00216 & 0.0230   & 0.0250   & 0.0210 & 0.0230 & 0.0050   & 0.0030   & 0.0148 & 0.0140 \\
223 &  5.237  & 0.00771  & 0.00736  & 0.00065 & 0.00761 & 0.0184   & 0.0184   & 0.0088 & 0.0200 & 0.0059   & 0.0059   & 0.0082 & 0.0102 \\
225 &  2.613  & 0.01061  & 0.01053  & 0.01032 & 0.00998 & 0.0344   & 0.0344   & 0.0311 & 0.0344 & 0.0073   & 0.0073   & 0.0224 & 0.0174 \\
233 &  3.083  & 0.00461  & 0.00460  & 0.00431 & 0.00459 & 0.0153   & 0.0153   & 0.0153 & 0.0153 & 0.0035   & 0.0035   & 0.0040 & 0.0049 \\
\hline                  
\label{tab:trpara}
\end{tabular}
\end{table*}

\begin{table*}
\caption{Star-planet parameters (planet to star radius ratio $R_{p}/R_{\star}$, system scale $a/R_{\star}$, and impact parameter $b$) derived from full transit fits. The columns corresponding to the 4 sets of light curves used in the fits are labelled as in Table~\ref{tab:trpara}.}             
\centering          
\begin{tabular}{l r c c c c r r r r c c c c} 
\hline\hline       
LC &  period & \multicolumn{4}{c}{$R_{p}/R_{\star}$} & \multicolumn{4}{c}{$a/R_{\star}$}& \multicolumn{4}{c}{$b$} \\
& (days) & no noise & no stvar  & NIF & IRF & no noise & no stvar  & NIF & IRF & no noise & no stvar & NIF & IRF \\
\hline 
\hline 
126    &  4.576   & 0.0799    & 0.0800 & 0.0698 & 0.0817   & 12.27     & 12.09     & 13.14 & 11.89    & 0.862	 & 0.870    & 0.858 & 0.872 \\
131    &  6.880   & 0.0760    & 0.1687 & 0.0779 & 0.0749   & 15.64     &  9.64     & 10.65 & 12.30    & 0.825	 & 1.058    & 0.893 & 0.873 \\
133    &  8.128   & 0.1389    & 0.2836 & 0.0557 & 0.0642   & 17.70     & 15.59     & 20.99 & 20.65    & 1.074	 & 1.233    & 0.961 & 0.979 \\
135    &  3.733   & 0.0469    & 0.0369 & 0.0372 & 0.0393   & 10.00     & 20.34     & 20.62 & 16.75    & 0.916	 & 0.422    & 0.397 & 0.676 \\
145    &  5.557   & 0.1050    & 0.1044 & 0.1080 & 0.1079   & 14.35     & 14.24     & 13.34 & 13.38    & 0.788	 & 0.795    & 0.822 & 0.820 \\
152    &  7.360   & 0.0481    & 0.0476 & 0.0495 & 0.0475   & 15.48     & 15.80     & 13.29 & 15.35    & 0.860	 & 0.836    & 0.908 & 0.865 \\
154    & 10.987   & 0.0263    & 0.0313 & 0.0323 & 0.0245   & 11.54     &  9.24     &  8.50 & 23.47    & 0.829	 & 0.914    & 0.933 & 0.023 \\
162    &  4.171   & 0.1335    & 0.1245 & 0.1102 & 0.1272   & 10.91     & 11.13     & 11.58 & 11.21    & 0.927	 & 0.910    & 0.911 & 0.913 \\
169    &  5.195   & 0.0918    & 0.0921 & 0.0781 & 0.0925   & 12.61     & 12.36     & 11.84 & 12.25    & 0.720	 & 0.736    & 0.750 & 0.735 \\
177    &  7.339   & 0.0486    & 0.0555 & 0.0549 & 0.0549   & 17.53     & 10.92     & 11.66 & 11.83    & 0.150	 & 0.794    & 0.760 & 0.749 \\
186    &  4.373   & 0.0839    & 0.0872 & 0.0869 & 0.0876   & 12.71     & 11.21     & 11.69 & 11.69    & 0.667	 & 0.760    & 0.738 & 0.735 \\
192    &  3.915   & 0.0504    & 0.0488 & 0.0152 & 0.0240   & 10.97     & 12.64     & 12.09 &  8.51    & 0.988	 & 0.978    & 0.656 & 0.926 \\
193    &  6.763   & 0.0904    & 0.0929 & 0.0990 & 0.0996   & 15.26     & 13.90     & 14.60 & 14.42    & 0.723	 & 0.780    & 0.759 & 0.764 \\
196    &  4.608   & 0.1162    & 0.1160 & 0.0970 & 0.1150   & 11.87     & 11.91     & 14.47 & 11.98    & 0.546	 & 0.533    & 0.000 & 0.547 \\
200    &  5.995   & 0.0860    & 0.3645 & 0.0444 & 0.0755   & 14.59     & 12.10     & 22.72 & 15.91    & 0.965	 & 1.290    & 0.916 & 0.945 \\
208    &  4.064   & 0.0582    & 0.0588 & 0.0597 & 0.0592   &  9.33     &  9.30     &  8.63 &  8.93    & 0.710	 & 0.716    & 0.780 & 0.763 \\
220    &  7.253   & 0.0475    & 0.0436 & 0.0433 & 0.0462   & 10.71     & 12.86     & 15.67 & 12.93    & 0.722	 & 0.519    & 0.019 & 0.550 \\
223    &  5.237   & 0.0976    & 0.0811 & 0.0452 & 0.1008   & 11.33     & 19.75     & 18.84 &  9.82    & 0.835	 & 0.017    & 0.231 & 0.854 \\
225    &  2.613   & 0.1033    & 0.1028 & 0.1017 & 0.0989   &  8.28     &  8.22     &  8.59 &  8.84    & 0.624	 & 0.625    & 0.575 & 0.552 \\
233    &  3.083   & 0.1426    & 0.1500 & 0.1414 & 0.1378   &  9.29     &  9.38     &  9.30 &  9.23    & 1.020	 & 1.029    & 1.020 & 1.012 \\
\hline 
\hline                  
\label{tab:plpara}
\end{tabular}
\end{table*}

\section{Contaminant flux}
\label{app:contam}

\begin{table}
\caption{Table of flux percentages coming from a contaminant star, for each of the BT2 light curve studied. Each light curves was corrected from the contaminant flux, before deriving transit and planet parameters. The fraction of flux coming from a contaminant star in each colour channel (CoRoT red, green and blue) was given in the parameter file used to build the BT2 light curves. For each light curve, the total contaminant flux was computed as the median of the sum of the contaminant fluxes in each colour channel, normalised by the median of the total flux.}             
\label{tab:contam}      
\centering          
\begin{tabular}{l c l c}     
\hline \hline       
LC &  contaminant flux (\%) & LC &  contaminant flux (\%)\\
\hline 
105 & 0.2 & 177 & 0.6 \\
110 & 0.1 & 186 & 0.3 \\
126 & 2.2 & 192 & 0.8 \\
131 & 90.6 & 193 & 13.1 \\
133 & 0.2 & 196 & 0.9 \\
135 & 0.1 & 200 & 3.3 \\
145 & 2.3 & 208 & 1.8 \\
152 & 0.3 & 220 & 1.9 \\
154 & 1.9 & 223 & 77.4 \\
162 & 0.1 & 225 & 0.6 \\
165 & 91.1 & 233 & 0.6 \\
169 & 0.5 & 236 & 1.4 \\
\hline \hline                  
\end{tabular}
\end{table}

\label{lastpage}

\end{document}